\documentclass[12pt,preprint]{aastex}

\usepackage{emulateapj5}
\usepackage{aastexug}
\usepackage{amsmath}
\usepackage{graphicx}
\usepackage{subfigure}
\usepackage{apjfonts}

\begin{document}
\def\gapprox{\mathrel{\vcenter{\offinterlineskip \hbox{$>$}
    \kern 0.3ex \hbox{$\sim$}}}}
\def\lapprox{\mathrel{\vcenter{\offinterlineskip \hbox{$<$}
    \kern 0.3ex \hbox{$\sim$}}}}

\title{Nonlinear Evolution of the Magnetohydrodynamic
Rayleigh-Taylor Instability}

\author{James M. Stone and Thomas Gardiner}
\affil{Department of Astrophysical Sciences, Princeton University, Princeton,
NJ 08544}

\begin{abstract}

We study the nonlinear evolution of the magnetic Rayleigh-Taylor
instability using three-dimensional MHD simulations.  We consider
the idealized case of two inviscid, perfectly conducting fluids of
constant density separated by a contact discontinuity perpendicular
to the effective gravity $g$, with a uniform magnetic field ${\bf B}$
parallel to the interface.  Modes parallel to the field with wavelengths
smaller than $\lambda_c = {\bf B}\cdot{\bf B}/(\rho_h - \rho_l)g$ are
suppressed (where $\rho_h$ and $\rho_l$ are the densities of the heavy
and light fluids respectively), whereas modes perpendicular to ${\bf B}$
are unaffected.  We study strong fields with $\lambda_c$ varying between
0.01 and 0.36 of the horizontal extent of the computational domain.
Even a weak field produces tension forces on small scales that are
significant enough to reduce shear (as measured by the distribution of
the amplitude of vorticity), which in turn reduces the mixing between
fluids, and increases the rate at which bubbles and finger are displaced
from the interface compared to the purely hydrodynamic case.  For strong
fields, the highly anisotropic nature of unstable modes produces ropes
and filaments.  However, at late time flow along field lines produces
large scale bubbles.  The kinetic and magnetic energies transverse to
gravity remain in rough equipartition and increase as $t^4$ at early
times.  The growth deviates from this form once the magnetic energy in
the vertical field becomes larger than the energy in the initial field.
We comment on the implications of our results to Z-pinch experiments,
and a variety of astrophysical systems.

\end{abstract}

\section{Introduction}

When a light fluid accelerates (or supports against gravity) a heavier
fluid, the interface between the two is subject to the Rayleigh-Taylor
instability (RTI).  The idealized case of two incompressible, inviscid,
unmagnetized fluids of constant density separated by a contact
discontinuity perpendicular to the effective gravity $g$ has been extensively
studied through theory$^{1}$, experiment$^{2,3}$ and numerical
simulation$^{3}$.  In the linear regime, short wavelength perturbations
of the interface grow fastest$^{1}$.  Once the perturbation amplitude
is comparable to the wavelength, the interface can be characterized
as rising ``bubbles" of light fluid between descending ``fingers"
of heavy fluid.  Secondary Kelvin-Helmholtz instabilities are induced
by the shear between the rising and descending columns.  In the fully
nonlinear phase of a multimode spectrum of perturbations, smaller
bubbles merge into larger structures, which then break up due to
secondary instabilities, and a turbulent mixing layer results.

From self-similarity arguments$^{3}$, the time evolution of the height
of the bubbles $h$ above the initial interface is expected to be
\begin{equation}
 h = \alpha Agt^2
\end{equation}
where $\alpha$ is a dimensionless constant, and $A$ is the Atwood number
\begin{equation}
A \equiv \frac{\rho_h - \rho_l}{\rho_h + \rho_l}
\end{equation}
($\rho_h$ and $\rho_l$ are the densities of the heavy and light
fluids respectively).  Recently, a comparison of the values of $\alpha$
measured from laboratory experiments with the results of high resolution,
three-dimensional numerical simulations of multimode RTI with strong
mode coupling, computed with a variety of numerical methods, has been
presented by Dimonte {\em et al}$^{3}$ (see also Ref. 4) as part of a
validation effort for numerical methods for hydrodynamics.  Detailed
analysis of the self-similar bubble dynamics, energy balance, and spectral
properties of the resulting turbulent mixing layer demonstrated there
is reasonable agreement between the simulations, theory and experiment,
except in that the experimentally determined value of $\alpha$ is $0.057
\pm 0.008$, whereas most of the numerical simulations give a value for
$\alpha$ which is about a factor of two smaller.  It would appear this
discrepancy is primarily due to mixing at small (close to the grid)
scales, since the use of specialized front tracking algorithms$^{4}$,
or correcting the numerically measured growth rates for the observed
density dilution produced by small scale mixing, or comparison to
experiments which use miscible fluids, all give better agreement.

It is important to validate numerical algorithms in experimentally
accessible parameter regimes (such as the hydrodynamic RTI) so that these
methods can be used with confidence to explore new physics in regimes
which may be hard to realize in the laboratory.  For example, there are
a number of applications where magnetic fields may play an important
role in the linear evolution and nonlinear saturation of the RTI.
The axial compression of plasma produced by the ablation of wires in
Z-pinch experiments$^{5}$ is subject to the magnetic RTI.  Since the
instability can limit the maximum compression achieved in the pinch,
understanding and controlling it is of critical importance.  Furthermore,
since most astrophysical plasmas are magnetized, the RTI associated
with accretion onto compact objects $^{6}$, supernova remnants$^{7}$,
magnetic flux emerging from the solar photosphere$^{8}$, and at the
surface of synchrotron nebulae expanding into supernova ejecta$^{9}$
is strongly influenced by the presence of magnetic fields.

For the ideal case introduced above, the linear growth rate $n$ of the RTI
with a uniform magnetic field of strength $B$ parallel to the
interface is given by$^{1}$
\begin{equation}
n^2 = gk \left( \frac{\rho_h - \rho_l}{\rho_h + \rho_l} 
  - \frac{({\bf B}\cdot{\bf k})^2}{2\pi gk (\rho_h + \rho_l)} \right)
\end{equation}
where instability occurs when $n^2 > 0$.  Here and throughout
we have chosen a system of units in
which the magnetic permeability $\mu=1$.  In the above, ${\bf k}$ is the
wavevector of a perturbation, and $k^2 = {\bf k} \cdot {\bf k}$.
For perturbations
perpendicular to the magnetic field the linear growth is identical
to the purely hydrodynamic case; these modes are often referred to
as interchange modes.  For perturbations parallel to the field,
wavelengths smaller than the critical value
$\lambda_c = 2\pi/k_c = B^{2}/(\rho_h - \rho_l)g$ are {\em stable}, and the
growth rate of modes at all scales larger than $\lambda_c$ is reduced
compared to the non-magnetic case.
Equivalently, for instability to occur on scales smaller
than $L$, the magnetic field must be smaller than the critical value
\begin{equation}
 B_c = [(\rho_h - \rho_l)gL]^{1/2}.
\end{equation}
Since the growth rate is zero at both large and small $k$, there
must be a wavelength $\lambda_{\rm max}$ where the growth rate is
maximum; it occurs at $\lambda_{\rm max} = 2\lambda_c$.  The presence
of a critical wavelength, a peak in the growth rate at $\lambda_{\rm
max}$, and the anisotropic nature of the dispersion relation for
perturbations parallel versus perpendicular to field lines suggests
the nonlinear evolution of the magnetic RTI will be much different
than the non-magnetic case.

Due to the anisotropic nature of the linear modes, it is critical
to study the magnetic RTI in full three dimensions.  Two-dimensional
ideal MHD studies in which the magnetic field is perpendicular to the domain
are in fact equivalent to 2D hydrodynamics with the gas pressure
$P$ replaced by the total pressure $P^* \equiv P + ({\bf B \cdot
B})/2$.  On the other hand, two-dimensional studies in which the
magnetic field is in the plane of the computation miss the interchange
modes, which artificially suppresses the instability.  For example,
in 2D with $B > B_c$ the interface is completely stable, whereas
in 3D it will be strongly unstable due to the interchange modes
(which will act like 2D hydrodynamic RTI in the plane perpendicular
to the field).

There have only been a few previous investigations of the magnetic
RTI in full three dimensions.  Jun et al.$^{10}$ presented
a few 3D simulations, although at much lower resolution than is
currently possible.  Their focus was in potential field amplification
and dynamo action produced by turbulent mixing in the saturated
state, thus most of their 3D simulations began with a field weak
compared to $B_c$ so that $\lambda_c$ was initially unresolved.
Zhu et al.$^{11}$ performed simulations of single mode interchange
instabilities in three dimensions at very low $\beta \equiv P/P_{\rm m}$,
where $P$ and $P_{\rm m}$ are the gas and magnetic pressures respectively,
with the goal of testing the predictions of perturbation theory on the 
nonlinear structure that emerges for a single mode.
More recently, Isobe et al.$^{8}$ showed that the magnetic RTI
can produce filamentary structures during the buoyant
rise of flux in the solar photosphere.

In this paper, we use numerical MHD simulations, using methods that
have been validated in the hydrodynamic regime, to study the nonlinear
evolution of the magnetic RTI with strong fields in three dimensions.
By strong we mean $\lambda_c$ is comparable to the size of the
computational domain.  Our goal is to explore how strong fields affect
the formation, structure, and evolution of bubbles and fingers in
the nonlinear regime, and how they affect the turbulent mixing layer
(and vice-versa).  Since we do not use front tracking methods, our
calculations are limited by the numerical diffusion that occurs near the
grid scale$^{3,4}$.  By comparison of hydrodynamic and MHD simulations
computed with the same parameters, numerical resolution, and algorithm,
we can assess the {\em relative} rate of mixing between the magnetic
and non-magnetic RTI.  Interestingly, we find that even with an initial
field too weak to resolve $\lambda_c$ (so that one might expect it to
evolve more like a hydrodynamic model), the mixing rate between the
light and heavy fluids is substantially reduced, and the rate of rise
of the bubbles as measured by the $\alpha$ parameter is substantially
increased with MHD.  Although such fields may be too weak to stabilize
resolved modes, they still add a significant ``surface tension" at
very small scales, which supports the idea that the discrepancy between
the experimentally measured value of $\alpha$ in the hydrodynamic RTI
and numerical simulations that lack front tracking is due to small scale
mixing.  Most of our calculations are motivated by the astrophysical
applications of the magnetic RTI, thus we study inviscid fluids in
the ideal MHD approximation (without any explicit resistivity) and in
a planar geometry.  The latter requires the thickness of the mixing
zone between the heavy and the light fluids be much smaller than the
radius of curvature of the interface $R$, and that $\lambda_c/R \ll 1$.
The magnetic RTI associated with Z-pinch experiments occurs at a very low
magnetic Reynolds number, and in a cylindrical geometry.  The inclusion
of non-ideal MHD effects$^{12}$, and the appropriate geometry, will be
important for future studies with application to Z-pinch$^{13}$.

The organization of this paper is as follows.  In the next section,
we describe the equations we solve, our numerical algorithm, and
the initial conditions.  In section 3 we present most of our results,
while in section 4 we summarize and discuss the application of our
results.

\section{Method}

We solve the equations of ideal MHD with a constant vertical
acceleration ${\bf g} = (0,0,-g)$ 
\begin{eqnarray}
\frac{\partial \rho}{\partial t} +
{\bf\nabla\cdot} \left(\rho{\bf v}\right) & = & 0
\label{eq:cons_mass} \\
\frac{\partial \rho {\bf v}}{\partial t} +
{\bf\nabla\cdot} \left(\rho{\bf vv} - {\bf BB}\right) +
{\bf \nabla} P^* & = & \rho {\bf g}  \\
\frac{\partial {\bf B}}{\partial t} +
{\bf\nabla} \times \left({\bf v} \times {\bf B}\right) & = & 0 \\
\frac{\partial E}{\partial t} +
\nabla\cdot((E + P^*) {\bf v} - {\bf B} ({\bf B \cdot v})) & = & \rho {\bf v}\cdot{\bf g}
\label{eq:cons_energy}
\end{eqnarray}
In these equations,
$P^*$ is the total pressure (gas plus magnetic), and $E$ is the total
energy density, which is related to the
internal energy density $\epsilon$ via
\begin{equation}
E \equiv \epsilon + \rho({\bf v \cdot v})/2 + ({\bf B \cdot B})/2 ~.
\end{equation}
We use an ideal gas equation of state for which $P =
(\gamma - 1) \epsilon$, where $\gamma$ is the ratio of specific heats.
We take $\gamma=5/3$.   Although we are solving the equations of
compressible gas dynamics, we choose a sound speed which is so large that
the resulting motions are highly subsonic and nearly incompressible.
Thus, varying the adiabatic index should have little effect on the
results reported here.

The calculations are performed in a three-dimensional domain of
size $L \times L \times 2L$, where $L=0.1$.  The vertical coordinate
spans $-0.1 \leq z \leq 0.1$.  The interface between heavy and light fluids
is initially at $z=0$.  The upper half of the domain ($z>0$)
is filled with heavy fluid of density $\rho_h=3$, while in the lower
half ($z<0$) the density of the light fluid is $\rho_l=1$.
The Atwood number is $A = 1/2$.  The profile
of the pressure is given by the condition of hydrostatic equilibrium,
while the amplitude is chosen so that the sound speed in the light
fluid is unity at the interface, thus
\begin{equation}
P^*(z) = \frac{3}{5} - g \rho z + B^2/2
\end{equation}
We choose $g=0.1$.  The sound crossing time $t_s$ in the light fluid
at the interface is 0.1.  We will report evolutionary times in our
computations in terms of $t_s$.  Periodic boundary conditions are
used in the transverse ($x-$ and $y-$) directions, and reflecting
boundary conditions are used at the top and bottom.

The magnetic field is initialized to be uniform and along the $x-$axis
${\bf B} = (B_0,0,0)$.  From the linear analysis (eq. [4]), if $B_0 >
B_c=0.14$, then there will be no unstable wavelengths shorter than
the size of the computational domain $L$.  We study the nonlinear
evolution for a variety of field strengths between 0.1 and $0.6B_c$.
For the strongest field $B_{0}=0.6B_c$, the ratio of the Alfv\'{e}n
speed to the sound speed in the light fluid at the interface is 0.024,
which corresponds to $\beta = 1.2(c_s/V_A)^2 \approx 2 \times 10^3$.
For the weakest field $B_{0}=0.1B_c$, $\beta = 7.5 \times 10^4$.  Thus,
we study a regime where the magnetic energy density is small compared to
thermal pressure (high $\beta$), and the vertical hydrostatic equilibrium
is determined by gas pressure alone.  Nonetheless, we study {\em strong}
fields in the sense that modes parallel to ${\bf B}$ are nearly completely
suppressed.  The magnetic RTI in plasmas with $\beta \approx 1$, which
is often referred to as the Parker instability in astrophysics$^{14}$,
has been extensively studied in the literature$^{15}$.

An important dimensionless parameter which characterized our simulations
is the ratio of the critical unstable wavelength to the size of the
computational domain $\lambda_{c}/L$.  For the strongest fields studied
here, $\lambda_{c}/L=0.36$.  Thus, only a few unstable modes are present
in the domain in this case.  This regime is appropriate to
physical systems in which some scale in the system (e.g. the diameter of
a wire in a Z-pinch) is comparable to the critical unstable wavelength.
In \S 4 we discuss the interpretation of our results in terms of this
parameter.

The RTI is seeded by small amplitude, random, zone-to-zone perturbations
to the vertical velocity $v_z$ added throughout the volume.  The
amplitude of the perturbations is smoothed toward the vertical
boundaries; $v_{z}(z) = A_0 R (1 + \cos{2\pi z/L})$ where $A_0=0.005$,
and $R$ is a random number between -1 and 1.  The peak perturbations are
therefore 1\% of $c_s$.  We have found that
perturbing the vertical velocity is superior to perturbing the
position of the interface, since the latter requires smoothing at
the grid scale when the perturbation amplitude is smaller than a
grid zone.  Dimonte {\em et al.}$^{3}$ were careful to introduce
a spectrum of multimode perturbations which is truncated at high
wavenumbers equivalent to 32 gridpoints, so that all linear modes
are initially resolved.  Instead, our perturbation spectrum is white
noise down to the grid scale.  This may mean that at very early
times, when modes are non-interacting, and the high-$k$ modes
dominate, there may be differences between our hydrodynamical
simulations.  However, since we follow the multimode evolution
deep into the nonlinear regime, where mode interactions should erase
memory of the initial conditions, we do not expect this to limit
comparisons at late times.

The computations presented in this paper are computed using a recently
developed Godunov method for compressible MHD that combines the
piecewise parabolic method$^{16}$ and the directionally unsplit corner
transport upwind (CTU) integrator$^{17}$ with the constrained
transport$^{18}$ algorithm for enforcing the divergence free constraint.
A complete description of the algorithm, including the results of an
extensive series of test problems, is given in Ref. [19].
Adding source terms (vertical gravity) to a Godunov scheme requires
particular care, by adding them to both the PPM reconstruction step,
as well as to the transverse flux differences in the CTU integrator,
we find our method holds the vertical equilibrium state (in which the
pressure gradient balances gravity) exactly.

When run in hydrodynamic mode (which utilizes the same algorithms as
when the code is configured for MHD, except for the Riemann solver),
the algorithm is similar to the FLASH and WP/PPM codes used by Dimonte
{\em et al}, except for the use of the unsplit CTU integrator.  All of
the simulations use a grid of $256 \times 256 \times 512$, the highest
resolution reported in Dimonte {\em et al}.  In fact, apart from the
perturbation spectrum, we use the same grid and parameters in order to
more easily facilitate comparisons.  We have tested for convergence of
our numerical solutions by running a few simulations at one half this
resolution ($128 \times 128 \times 256$).  Although such details as the
number and location of bubbles and fingers in the flow are substantially
different at lower resolution, none of the diagnostics which are the
focus of this work are changed to a significant degree.  This indicates
our solutions are converged with respect to these quantities.
We provide a much more comprehensive investigation of the effect of mass
diffusion at the grid scale due to numerical effects on
our results in \S 3.1.

The use of a fully compressible code to study low Mach number flows such
as investigated here is not optimal, and requires many more timesteps
to be taken compared to the strictly incompressible case (typically
about $4\times10^{4}$ timesteps are required to reach $t/t_{s}=60$).
The numerical algorithms used here are second order in both space
and time$^{20}$, which helps to reduce the accumulation of error.
The convergence study presented in \S 3.1 provides insight into the
effect of temporal errors on our results.  Although approximate methods
for low Mach number flows might be more cost effective, it is not clear
they will be substantially more accurate.

\section{Results}

We describe the results from four MHD simulations using field
strengths of $B_0 =$ 0.1, 0.2, 0.4, and 0.6 $B_c$.  We shall refer
to each of these calculations as runs R1, R2, R4, and R6 respectively.
For comparison purposes, we also describe the results of a hydrodynamic
calculation computed with the same code, hereafter referred to as run RH.
Table 1 lists important properties of the calculations.  In particular,
note that for the weakest field simulation, R1, the critical wavelength
$\lambda_c$ is only about 2.5 grid cells, therefore it is unresolved.
For the smallest well resolved modes (requiring about 16 grid cells per
wavelength), the growth rate is reduced by only about 15\% at this field
strength compared to the non-magnetic case.  Thus, we expect run R1 to be
similar to the purely hydrodynamic calculation RH.  Our discussion will
focus on the strong field run R6, the weak field run R2 (the weakest
field for which the critical wavelength $\lambda_c$ is resolved),
and the hydrodynamic run RH.  Each calculation is continued until the
rising bubbles or sinking fingers reach the vertical boundaries, which
typically requires about $60t_s$.

\subsection{Convergence Study in Two Dimensions}

Before presenting the complex nonlinear evolution of the magnetic RTI
in fully three dimensions, it is important to first assess the extent
to which mixing of mass and momentum due to numerical effects at the
grid scale affect our results.  We have used a series of simulations
of the growth of the magnetic RTI in two dimensions, computed with
the identical algorithms and using the same parameters as used for the
three-dimensional runs listed in Table 1, to investigate the convergence of
our results.  The calculations are performed in the $x-z$ plane, so that
the magnetic field is in the plane of the computation.  We perturb the
interface with a single mode with a wavelength equal to the horizontal
size of the domain $L$, to investigate how numerical grid effects alter the
evolution of a single, smooth interface.

To track the amount of mixing between the heavy and light fluids,
we define a mixing parameter
\begin{equation}
\Theta = 4 f_{h} f_{l}
\end{equation}
where $f_h$ is the fraction of each grid cell occupied by the heavy fluid
(of density $\rho_h$), and $f_{l} = 1-f_{h}$.  For a purely incompressible
flow, with $\rho_h=3$ and $\rho_l=1$, then $f_h = (\rho -1)/2$.
In regions with no mixing, $\Theta=0$, while the
maximum value occurs for $\langle f_h \rangle = \langle f_l \rangle =1/2$,
when $\Theta=1$.  A useful diagnostic is the volume averaged mixing
parameter, which in two dimensions is
\begin{equation}
\langle \Theta \rangle_{V} = \int_{x}\int_{z} 4 f_{h} f_{l} dx dz / 2L^{2}
\end{equation}
A comparison of the time evolution of $\langle \Theta \rangle_{V}$
measured in our simulations with simple analytic expectations allows us
to measure the effect of mixing at the grid scale.

Figure 1 presents images of the mixing parameter in the evolution
of the single mode RTI in two dimensions using a strong field (run R6)
at time $t/t_s = 30$ for four different resolutions corresponding to 32,
64, 128, and 256 grid points per $L$.  Note the highest resolution is the
same as is used in all of the three-dimensional results presented in this paper.
The images show that $\Theta$ is non-zero only at the interface, which is
smooth and has the same shape at every resolution.  Clearly, the growth
rate and nonlinear structure of a single mode is captured correctly even
at the lowest resolution.  Although the physical width of the interface
over which mass mixing occurs is larger at low resolution, this is
simply due to the increase in the size of grid cells $\delta x = \delta z$;
it has not affected the rate of growth or structure of the mode.
Figure 1 gives the visual impression that the amount of mass diffusion at
the grid scale converges at first order (linear in $\delta x$),
we quantify this dependence below.

The initial conditions used in all of our simulations consist of a density
(and temperature) discontinuity at the interface between the heavy and
light fluids.  Initially this discontinuity is aligned with the grid.
For the numerical algorithms used in this paper (Godunov method with a
Roe solver), there is no numerical diffusion of this discontinuity as
long as it remains at rest parallel to the grid.  (We have confirmed
that if the interface is not perturbed, our code holds the initial
equilibrium state, and the mixing parameter remains $\Theta=0$ everywhere.)
Once the interface is perturbed and the RTI begins to grow, however,
motion of the interface produces mass diffusion at the grid scale.
This is an inevitable consequence of using a control volume approach
without interface tracking: when the interface crosses the middle of
the cell, the volume averaged density contains contributions from both
the heavy and light fluids, and will therefore be intermediate between
the two.  Figure 1 demonstrates that for the algorithms used here, the
spread of the contact discontinuity is confined to a few cells (several
$\delta x$),
Hence, at early times (when the length of the interface is just $L_{x}$)
we expect that 
\begin{equation}
 \langle \Theta \rangle_{V} \sim \frac{L_x M
 \delta z}{L_x L_z} \tilde{\Theta} =
 \frac{M \tilde{\Theta}}{N_{z}}
\end{equation}
where $M$ is the number of cells over which the interface spreads due
to numerical effcts,
$\tilde{\Theta}$ is the normalized average of $\Theta$ over the
width of the mixing region, and $N_{z}$ is the number of grid cells
in $L_{z}$.  If the variation of the heavy and light fluid fractions is
linear over the width of the mixing region, then $\tilde{\Theta} = 2/3$.
As the RTI grow, the length of the interface grows (e.g. figure 1).  Thus,
equation 13 expresses the expectation that the time evolution of $\langle
\Theta \rangle_{V}$ should be proportional to the time evolution of the
length of the interface, and that at any time $\langle \Theta \rangle_{V}$
should converge with $\delta x$ at first order (as expected for any
discontinuous solution using a fixed grid).

Figure 2 plots the time evolution of $\langle \Theta \rangle_{V}$ for
the single mode RTI in two dimensions (run R6) at the same resolutions
shown in figure 1.  There is a rapid rise in $\langle \Theta \rangle_{V}$
to $t/t_{s}=5$, as the interface begins to move across the grid, and
numerical diffusion causes it to spread to a size of a few $\delta
x$.  The amplitude of $\langle \Theta \rangle_{V}$ at $t/t_{s}=5$ is in
excellent quantitative agreement with the expectation of equation 13,
if the width of the interface $M \approx 2$.  Thereafter, $\langle
\Theta \rangle_{V}$ shows slow linear growth as the interface lengthens.
At $t/t_{s}=20$, the mode begins to go nonlinear, the length of the
interface begins to grow more rapidly, and $\langle \Theta \rangle_{V}$
increases more rapidly.  The fractional rate of increase $(1/\langle
\Theta \rangle_{V}) d \langle \Theta \rangle_{V}/dt$ is the same in
each case, indicating the increase in time is simply due to the lengthening
of the interface.  At $t/t_{s} = 20$, the convergence rate of $\langle
\Theta \rangle_{V}$ is 0.8, close to our expectation of first order.

The analysis presented thus far has considered an interface which remains
smooth, unaffected by secondary KH instabilities which are present with
weaker magnetic fields.  By twisting the interface on small scales, these
secondary instabilities can increase the mass mixing at the grid scale.
Figure 3 presents images of the mixing parameter $\Theta$ at 
$t/t_{s}=30$ in the evolution of a single
mode RTI computed with a resolution of 256 grid points per $L$, but with
a variety of field strengths corresponding to runs R6 (strong field), R2
(intermediate field), R1 (weak field), and RH (hydrodynamic); see Table 1.
The growth of an increasingly larger number of vortices at the interface is
evident as the magnetic field strength is decreased.  As these vortices wind
up the interface, mass mixing occurs due to numerical effects when the 
radius of curvature approaches the grid scale.

Figure 4 plots the time evolution of the volume averaged mixing parameter
$\langle \Theta \rangle_{V}$ for these runs.  Interestingly, until the 
appearance of the first KH roll at $t/t_{s}=20$, the amplitude and
evolution of $\langle \Theta \rangle_{V}$ is identical.  Thereafter,
the values of $\langle \Theta \rangle_{V}$ increase rapidly in runs R2, R1,
and RH, reflecting the increased importance of the small scale distortion of
the interface due to KH instabilities.

The fact that the time evolution of $\langle \Theta \rangle_{V}$
is identical regardless of the field strength for $t/t_{s}<20$ (until
the first KH roll forms) indicates the intrinsic spread of a smooth
interface in our numerical methods is independent of the field strength.
This gives us confidence that the relative amount of mixing we observe
in three dimensions between hydrodynamic and MHD simulations with
different field strengths will be due to physical differences in the
amount of geometric distortion of the interface, rather than a change
in the intrinsic numerical mixing inherent in the algorithm between
hydrodynamics and MHD.  Moreover, increasing the resolution will not
reduce this mixing, rather it will simply introduce more small scale structure
that produces a similar amount of mixing, unless a physical process such as
surface tension or viscosity is introduced to create a scale on which such
motions are suppressed.  We conclude the {\em relative} rate of mass mixing
we observe between hydrodynamic and MHD is robust.

The analysis above has focused on quantifying the amount of mass mixing
due to numerical effects in our algorithms.  Since we do not include
an explicit viscosity, momentum diffusion at the grid scale is also
dominated by grid effects.  For the algorithms used here, it has been
shown$^{20}$ that the properties of the numerical diffusion of momentum
produces proper convergence to a solution of the Navier-Stokes equations.
The effective Reynolds number of the flow is determined by the grid
resolution.  Since we adopt the same resolution for all runs,
the simulations reported here are equivalent to a study of the change
in the flow due to the effect of the magnetic field at fixed Reynolds number. 
Comparison to a specific experiment$^{21}$, rather than comparative results
as reported here, would require simulations that
achieve the same Reynolds number as the experiment.

Finally, the analysis presented above has focused on the evolution of
the RTI in two dimensions.  In fully three dimensions, the value of the
volume averaged mixing parameter $\langle \Theta \rangle_{V}$ will be
given by equation 13 with the length of the interface replaced by the
surface area.   The time evolution of $\langle \Theta \rangle_{V}$will
then depend on the rate of growth of this surface area, which will be more
rapid than in two dimensions.

\subsection{Three-dimensional structure}

Figure 5 is a comparison of the three-dimensional structure at both
early ($t/t_s=29.6$) and late ($t/t_s=60$) times from runs RH, R2, and R6
using vertical slices of the density at the edges of the computational
domain, as well as a horizontal slice at the midplane $z=0$.   The
hydrodynamic calculation RH (top two panels) can be compared directly
to the results in Ref. [3].  The classic evolution of the hydrodynamic
RTI is evident; at early times most structure is at small scales.
Bubbles of light fluid have detached from the interface, and are
dominated by a ``mushroom cap" appearance.  At late times, the
dominant structures are on larger scales.   Secondary Kelvin-Helmholtz
instabilities have produced vortices and mixing along the edges of
the bubbles, and a fully developed turbulent mixing layer is evident.

In the weak field calculation R2 (middle panels), the critical
wavelength at which the magnetic field can suppress the RTI is
small, only $0.01L$,
much smaller than the largest
bubbles seen at late times in RH.  Thus, we might expect the structure
produced in the nonlinear regime in R2 to be similar to the
hydrodynamic case RH.  At early times, R2 does show the classic
bubble and finger morphology of the hydrodynamic RTI.  There is no
evidence for anisotropy; structures are identical parallel and
perpendicular to the field.  However, even at the early time there
is clearly much less mixing between the light and heavy fluids; the
horizontal slice shows that at the midplane most of the fluid is either
at the highest or lowest density, whereas in RH most of the fluid is at
intermediate (grey) densities (for online version in color, high density
is red, low density is blue, intermediate density is green).  At late
times, the structure of the RTI is much different in R2 in comparison to
RH.  Rather than a turbulent mixing layer, in R2 the bubbles have become
long columns which have extended far above the midplane.  These structures
show no anisotropy.  There continues to be little mixing between the
fluids,  Thus, even a weak field has strongly affected the structure.

In the strong field calculation R6 (bottom panels) the anisotropic
nature of linear modes is clearly evident: the density fluctuations
at the midplane are aligned with the direction of the field (along
the $x-$axis).  As in R2, very little mixing is evident at the
midplane in R6.  At late times in the strong field case, the dominant
structures are smooth and highly anisotropic.  The slice in the $y-z$
plane shows columns and bubbles which result from interchange modes
that act like the hydrodynamic RTI in two dimensions.  In the $x-z$
plane, however, extended loop-like structures are evident which
result from suppression of short wavelength modes along the field.
These structures are unlike any of the bubbles seen in the hydrodynamic
RTI (run RH, top panel).

To further illustrate the nonlinear structure at late time,
Figure 6 shows isosurfaces of the density at $\rho=1.1$ and 2.9, along
with vertical slices of the density at the faces of the computation domain
for runs RH and R2 at $t/t_s=56$.  A complex network of bubbles is evident
in the shape of the density isosurface in RH, whereas in R2 the bubbles
are much larger and smoother.  Note the circular ring near the rear edge of
the domain in RH resulting from the roll-up of the KH instability around a
spherical bubble.  The slices at the edge of the domain also make clear the
turbulent mixing in RH, whereas in R2 there is far less mix.

Isosurfaces and slices of the density in the strong field run R6
are shown at two times in Figure 7; the left panel is at $t/t_s=20$ and
the right at $t/t_s=56$.  At the earlier time, the isosurface reveals
the formation of filaments and tubes rather than bubbles and fingers
as in the hydrodynamic case.  The anisotropy of linear modes is
evident in the comparison of the slices in the $x-z$ and $y-z$
planes.  In the former, the magnetic field has suppressed short
wavelengths, and the density structures are highly elongated in the
$x-$direction.  In the latter, the magnetic field is perpendicular
to the plane, and thus has no effect.  Bubbles and fingers on short
wavelengths have formed, reminiscent of the 2D hydrodynamic RTI.
The formation of flux tubes seen at early times is very similar to
that observed in the solar photosphere, as shown in Ref. [8].  At
late times (right panel of Figure 7), the density isosurface is
remarkably smooth.  Rather than sheets, more rounded bubbles have
been formed at the tips of the columns by the flow of fluid along
loops until it collects at the tips.  Similar evolution is observed
in the nonlinear regime of the Parker instability$^{15}$.

An important diagnostic of the instability is the rate at which 
bubbles and fingers are displaced from the initial interface.  To 
quantify this rate, we define the horizontally 
averaged mass fraction of heavy fluid as (following Ref. [3])
\begin{equation}
 \langle f_h \rangle \equiv \int_x \int_y f_h dx dy/L^2.
\end{equation}
Since our
simulations are not purely incompressible, the maximum
(minimum) densities can be slightly larger (smaller) than 3 (one).  To
account for this, we define the height of bubbles to be the location
where $\langle f_h \rangle = 0.985$, while the height of fingers
is the location where $\langle f_h \rangle = 0.015$.

Figure 8 is a plot of the height of bubbles and fingers in runs RH,
R2, and R6 versus $t^2$.  From the self-similar arguments that lead to
eq. [1], we expect at late time a straight line with a slope of $\alpha$.
In each case, this expectation is confirmed.   In the hydrodynamic run
RH, the best fit slope at late time is $\alpha = 0.021$, whereas for
MHD we obtain $\alpha=0.035$ for both R4 and R6.  Although not plotted to avoid
cluttering the figure,
we have also confirmed that for the weakest field run R1, the height of
bubbles follows a straight line when plotted versus $t^2$, with a slope
of $\alpha=0.031$.  Thus, we find that (1) the rate at which bubbles
rise in the hydrodynamic RTI measured in our simulations agrees with
the results of Dimonte et al for non-front tracking methods, and (2) the
addition of even a small magnetic field significantly increases the slope.
We show below that there is far less mixing in the MHD RTI simulations,
which may account for this increase.

\subsection{Mixing}

Figure 9 plots the vertical profile of the fraction of heavy fluid
$\langle f_h \rangle$ for runs RH and R6 at $t/t_s=56$, where the
horizontal axis has been scaled by the bubble height at that time.
The profiles are remarkable similar to each other.  We also find the
profiles are the same at different times in the evolution, as is expected
due to the self-similar evolution.  The similarities in the profiles
of $\langle f_h \rangle$ between the magnetized and unmagnetized RTI,
despite the lack of a turbulent mixing layer in the former, suggests
that this profile is not sensitive to mixing.

Figure 10 plots the vertical profile of the horizontally averaged
mixing parameter 
\begin{equation}
\langle \Theta \rangle =4\langle f_h f_l \rangle
\end{equation}
for
runs RH, R1, R2 and R6 at $t/t_s=56$.  There is a monotonic decrease in
the maximum value of $\langle \Theta \rangle$ with the field strength.  The hydrodynamic
case RH is closest to reaching $\langle \Theta \rangle=1$, the maximum value possible.
Even the weakest field case, run R1, in which the critical wavelength
suppressed by the magnetic field is unresolved initially, has a 
significantly lower value of $\langle \Theta \rangle$, indicating far less mixing is
occurring compared to the hydrodynamic case.

It also is of interest to compare the integral of the mixing parameter
over the vertical coordinate between different runs.  For the
hydrodynamic and weak field runs RH and R1 respectively, $\int \langle \Theta \rangle
dz \approx 0.06$, with smaller values obtained for the stronger field
cases R2 and R6.  This quantity has dimensions of length and so can be
interpreted as the effective width of a completely mixed region (for
which $\langle 4 f_l f_h \rangle = 1$).  The volume averaged value
$\langle \Theta \rangle_{V} = \int \langle \Theta \rangle dz /L \approx 0.6$, which
can be interpreted as the fractional height of the domain over which
mixing occurs.  Thus, not only is the mixing
local to the initial interface larger for hydrodynamics compared to MHD,
but also even in an integral sense the effective width is reduced.

We have found the variance of the density is also a sensitive
diagnostic of the mixing region.  At each vertical position $z$,
we compute the horizontally averaged density $\langle \rho \rangle$,
where the $\langle \rangle$ denotes an average over a horizontal
plane as in eq. 11.  The variance is then
\begin{equation}
\delta \rho(z) = \langle (\rho - \langle \rho \rangle)^2 \rangle^{1/2}
\end{equation}
Regions which are well mixed have a smaller variance.  If the heavy
and light fluids remain completely unmixed (that is, if the density
at every location can only be either 3 or one) then the largest
value of the variance is $\delta \rho/\langle \rho \rangle =
1/\sqrt{3} \approx 0.57$ and occurs for $\langle f_h \rangle =
1/4$.

Figure 11 plots the vertical profile of $\delta \rho/\langle \rho
\rangle$ in Runs RH, R2, and R6 at $t/t_s=56$.  Both the weak and
strong field runs are similar to each other, and have a much larger
variance than the unmagnetized case, indicating less mixing.  For
$\langle f_h \rangle =1/2$, which from a visual inspections of the
horizontal slice plane in Figure 5 is a rough approximation to the
volume fraction of the heavy fluid there, the ideal case of no
mixing gives $\delta \rho/\langle \rho \rangle =1/2$.  From Figure
11 the peak values in runs R2 and R6 approach this value, an indication
that very little mix occurs in the MHD RTI.  Note from Figure 9
that $\langle f_h \rangle = 1/4$ occurs at $Z/H$ of -0.4.  At the
time of this plot, the height of the bubbles $H \approx 0.7$, thus
we should expect the peak in the variance to occur at $z \approx
-0.03$, which is in good agreement with Figure 11.  Thus, the asymmetry
in the variance (with larger values occurring for negative $z$) is a
consequence of the lower $\langle f_h \rangle$ there.

To investigate the dynamical processes that can lead to reduced
mixing in the MHD RTI compared to hydrodynamics, we have calculated
the distribution of the amplitude of the vorticity $\mid W \mid = \mid
\nabla \times {\bf v} \mid$.  Figure 12 plots the number of cells $N$ in
runs RH and R1 which have a given value of the amplitude of vorticity,
$N(\mid W \mid )$, versus $\mid W\mid$.  Clearly there is a shift in the
distribution to lower values of $\mid W \mid$ in the weak field run R1
(dotted line).  Large values of the vorticity are associated with velocity
shear over small scales.  The shift in the distribution to lower $\mid
W \mid$ with weak fields is an indication that even if the critical
wavelength $\lambda_c$ is not resolved in the initial conditions
(R1), the tension forces associated with small scale bending of the
field lines is sufficient to affect the flow, and suppress the shear.
We postulate that the suppression of small scale shearing motion implied
by the change in the distribution $N(\mid W \mid )$ is responsible for
the reduced mixing observed in the MHD RTI as shown by figures 10 and 11.

\subsection{Evolution of the magnetic field}

To investigate the three-dimensional structure of the magnetic field,
Figure 13 plots slices of the magnetic energy in the fluctuating part of
the field,
\begin{equation}
\delta B^2/2 = (B_x - B_0)^2/2 + B_y^2/2 + B_z^2/2
\end{equation}
at the edges of the domain, and at the midplane $z=0$ in runs R2 and R6
at $t/t_s=60$.  The slices are made at the same locations and the same
times as the slices of the density shown in the right hand panels of
figure 5.  There is a direct correspondence to the structures visible
in the magnetic field and the density.  The fluctuations in $\delta B^2$
occur at smaller scales in R2 in comparison to R6.  Most of the magnetic
energy is concentrated in rope and sheet-like structures associated with
the bubbles and fingers in the density.  The maximum of the magnetic
energy is quite similar in the two plots (the maximum is 0.021 in R2
and 0.015 in R6), despite the fact the energy in the background field is
nearly an order of magnitude larger in R6.  In fact, at $t/t_s=60$, the
energy in fluctuations is larger in the weaker field run.  As discussed
below, this is likely due to the fact that the fingers and bubbles
are larger in R2 at this time, thus more gravitational energy has been
released that can be tapped to amplify the magnetic field.

Figure 14 is a plot of the the vertical profile of the horizontally
averaged magnetic energy at two times in both runs R2 and R6.  The energy
in the vertical component $\langle B_z^2/2 \rangle$ is separated from
the horizontal components $\langle (B_x - B_0)^2/2 + B_y^2/2 \rangle$.
The energy in the vertical field always dominates, a consequence of
the dominance of vertical motions.   The amplitude of the energy in the
vertical field is remarkably similar between the two runs; at $t/t_s =
56$ it is about 0.009 in R2 and 0.0075 in R6.  This is consistent with the
amplitude of the energies being similar in the slices shown in figure 13.
The total energy in fluctuations (the integral over vertical height of
the lines plotted in figure 14) must therefore be similar between the
two runs; we discuss this further below.  The ratio of energy in the
vertical versus horizontal fields is larger in the weak field case R2.

Simple energy arguments can be used to interpret the amplitude of the
magnetic energies observed in figures 13 and 14.  The gravitational potential
energy released by descending heavy fluid is converted into kinetic
energy by the RTI and secondary KH instabilities.  In turn, these result
in twisting and amplification of the magnetic field.  Some kinetic and
magnetic energy is converted into internal energy by viscous and resistive
dissipation.  Although our simulations do not include explicit dissipation,
our method conserves total energy exactly so that whatever energy
is lost by numerical viscosity and reconnection is explicitly captured
as an increase in internal energy.  The simplest expectation is that
the kinetic and magnetic energies will remain in approximate equipartition,
and that the amount of energy released when the tip of the fingers reaches
the same height (as measured by the vertical location where
$\langle f_h \rangle = 0.995$) will be the same amongst all runs.  To test
these ideas, table 2 compares the volume averaged energies from all
the runs when the tips of the fingers reaches $z/H=0.5$.  The third
column gives the time at which this happens, the fourth is the magnetic
energy in fluctuations, the fifth is the kinetic energy, the sixth is the
change in internal energy at that time.  The seventh column is the total
change in energy, and the last is the amplification factor of the magnetic
energy in fluctuations.  Since the data is collected not at the same
physical time, but at the point where the fingers have reached the same
height (and therefore the same amount of gravitational energy has been 
released), then the total of the kinetic, magnetic, and thermal energies should
be the same amongst all the runs.  Indeed, this expectation is confirmed,
the difference in total energy is only 15\% amongst all the simulations.
Note also the magnetic and kinetic energies are in rough equipartition,
regardless of the initial field strength.  The increase in internal energy
is a non-negligible contribution to the total change, especially in the
weaker field simulations.  In fact the
total gravitational potential energy released depends on the detailed
distribution of density in the interface region and not just the location
of the tip of the fingers, thus we do not expect the total energy
to be identical between the different runs.

These same energy arguments can be used to predict the time evolution of
the magnetic and kinetic energies.  The gravitational energy released
depends on the total mass and the distance it drops, both of which are
proportional to the height of the bubbles $h$.  Thus, the energy released
$E \propto h^2 \propto t^4$.  Figure 15 plots each component of the
magnetic and kinetic energies in runs R1, R2, R4, and R6 versus $t^4$.
Our expectation is that each should be a straight line, with rough
equipartition between the transverse magnetic and kinetic energies, and
with the vertical components dominating.  This expectation is clearly
borne out by the figure, at least so long as the energy in the vertical
component of the field $B_z^{2}/2$ (which grows the fastest) is less than
the magnetic energy associated with the initial field $B_{0}^{2}/2$.
Thus, during the evolution of strong field case R6 (figure 15a), the
ratio $B_z^{2}/B_{0}^{2}$ is always less than one, and the growth of
each component of the energy closely follows $t^{4}$.  With increasingly
weaker fields, the amplification of $B_z^{2}/B_{0}^{2}$ becomes larger
and larger, and the time evolution of each component diverges farther
and farther from $t^{4}$.  For the very weak field case, R1 (figure 15d),
the ratio $B_z^{2}/B_{0}^{2}=1$ is reached very early in the evolution,
and the maximum amplification at late time is more than 15.  In this case,
the time evolution of the energy is lower than $t^{4}$.
As discussed in the next section, we expect deviation from the simple $t^{4}$
scaling when the bubbles and fingers have moved a distance much larger then
$\lambda_c$.

\section{Summary and Discussion}

We have studied the RTI in ideal MHD and full three dimensions.  We have
restricted our analysis to uniform fields parallel to the interface.
We study the high-$\beta$ regime, where the energy density in the magnetic
field is small compared to the thermal energy in the fluid.  Nonetheless,
we study strong fields in the sense that the initial field strength $B_0
\lapprox B_c$, where $B_c$ is the critical field strength at which all
modes parallel to ${\bf B}$ are completely suppressed.  We use numerical
methods that have been validated in the sense that they reproduce the
growth rate of fingers and the amount of mixing between light and heavy
fluids in the hydrodynamic RTI, as reported in previous high resolution
numerical simulations$^{3}$, and laboratory experiment$^{2,3}$.

Uniform magnetic fields do not suppress the RTI, but rather make
modes strongly anisotropic.  Along the field, the growth rate of
modes is reduced, and wavelengths below a critical value are stable.
Perpendicular to the field, the dispersion relation of the magnetic
RTI is identical to the dispersion relation in hydrodynamics.  Even if
the field is made arbitrarily strong to suppress all modes parallel to
B, the density interface will still be RTI unstable in 3D due to the
interchange modes perpendicular to B.  It is therefore critical to study
the MHD RTI in full three dimensions.

Although uniform magnetic fields can not suppress the RTI in
three dimensions, they significantly change the nonlinear evolution
and structure.  Even weak fields reduce the mix between light and
heavy fluids, resulting in fingers and bubbles which rise much more
quickly compared to the hydrodynamics case.   Such fields are weak in
the sense that $\lambda_c/L$ is small, however they still are strong
enough to influence the flow through tension forces at small scales,
as evidenced by the change in distribution of the vorticity between
very weak field, and purely hydrodynamical, simulations (figure 12).
A turbulence mixing zone is not produced with strong fields.  Instead,
at early times the bubbles and fingers are elongated along ${\bf B}$,
forming flux ropes and tubes.  Fluid drains along these tubes, pooling
as bubbles at the tips and valleys, eventually forming the usual bubble
and finger morphology.  Interchange instabilities wrinkle the surfaces of
the bubbles perpendicular to the field.  Several diagnostics, including
the variance of the density, that are good diagnostics of the amount
of mixing between heavy and light fluids show that there is a monotonic
decrease in mixing with increasing field strength.

The evolution of the MHD RTI follows the same self-similar evolution
as in hydrodynamics.  The vertical profile of the volume fraction of
heavy fluid is self-similar.  The vertical extent of the bubbles and
fingers $h \propto t^2$.  Simple energy arguments suggest the magnetic
and kinetic energies should grow in time as $t^4$, and our results
confirm this expectation.  The energy in transverse motions and field
are always in rough equipartition.  The total energy in fluctuations
is independent of the initial field strength, but depends only on the
extent of the bubbles and fingers (which in turn determines the amount of
gravitational energy released that is available for field amplification).
Since the magnetic energy density in the mixing layer
grows to the same value in every case (determined by
the amount of gravitational binding energy released by descending
heavy fluid and the thickness of the mixing layer), this leads to a much larger fractional increase in magnetic
energy for initially weak fields (more than a factor of 10).  However,
this dynamo action can never lead to strong fields with $\beta \lapprox
1$, since the total magnetic energy is
limited to equipartition with
the kinetic rather than thermal energy, and the flows induced by the
RTI are highly subsonic.  One explanation for the larger increase in
magnetic energy with initially weak fields is that the critical
wavelength at which the RTI is stable is smaller for weak fields, thus
weak fields can be folded, twisted, and amplified on smaller scales than
strong fields.

The self-similar evolution of the magnetic and kinetic energies
diverges from the simple expectation above once the magnetic energy in
the vertical component of the field exceeds that associated with the
original field, i.e. $B_{z}^{2}/B_{0}^2 \geq 1$.  This occurs when the
bubbles and fingers have moved a large distance compared to $\lambda_c$.
The evolution of the energies in the strong field simulations described
here (R6), which has $\lambda_c/L=0.36$, closely follows a $t^{4}$ scaling
throughout.  However, for the very weak field simulation (R1), which has
$\lambda_c/L=0.01$, this scaling is broken at early times.  This difference
can be interpreted as being due to the very different dimensionless
length and time scales, $\lambda_c/L$ and $\sqrt{\lambda_c/g}/t_{s}$
respectively, in each case.  If the very weak field case R1 were repeated
with identical parameters but in a computational domain of size $L/36$,
then it would have the same ratio $\lambda_c/L$, and it would evolve
identically to the strong field case R6 on a time scale which is 6
times shorter.  Similarily, if the strong field case R6 were repeated
in a computational domain 36 times larger, it would be identical to
run R1 at times a factor of 6 longer than R1.  Thus, run R1 samples a
much later stage of evolution of the magnetic RTI than R6.  We conclude
the self-similar evolution which predicts energy growth at $t^{4}$ is
only applicable for $t \leq \tau_c = \sqrt{\lambda_c/g}$.
Although we have reported simulations of the magnetic RTI with
different field strengths and the same sized computational domain $L$
in this paper, this is equivalent to simulations with the same field
strength but with different sized computational domains.

The fact that the saturated magnetic energy is independent of the initial
field strength provides a simple explanation for the break in the $t^{4}$
scaling of the magnetic energy at late times $t>\tau_{c}$.  In this case,
since the magnetic field strength and magnetic energy density $B^{2}$
reaches a constant value in the mixing layer, then the total energy will
increase further only because of the thickening of the mixing layer,
thus in the saturated regime $E \propto h \propto t^{2}$.

The MHD RTI is relevant to a number of astrophysical systems, for example
density interfaces in the ISM, to the formation of filaments in the
Crab nebulae$^{9}$, and to the contact discontinuity in supernovae
blast waves$^{7}$.  As discussed above, a uniform field will not
suppress the RTI in these systems, nor even reduce the growth rate of
modes perpendicular to ${\bf B}$.  It will, however, result in highly
anisotropic structures, and reduce the mixing between fluids.  As reported
elsewhere (Stone \& Gardiner 2007, submitted to the Astrophys. J.), we
have also found that rotation of the direction of the field with vertical
position (which is equivalent to currents parallel to the interface)
significantly affects the nonlinear evolution of the the MHD RTI.

The MHD RTI is also relevant to Z-pinch experiments, where implosion
is driven by low density, highly magnetized plasma ablated from wires.
Non-ideal MHD effects (resistivity, and Hall currents) are important
at the densities and temperatures realized in the plasma in these
experiments$^{5}$.  Studies of the MHD RTI including resistive dissipation and
Hall currents in a cylindrical geometry with higher $\beta$ are needed$^{13}$.

\acknowledgements
We thank John Hawley for useful discussions, and for assistance in running 
the simulations, and the anonymous referees for suggestions that improved
the paper.  Simulations were performed on
the Teragrid cluster at NCSA, the IBM Blue Gene at Princeton University, and
on computational facilities supported by NSF grant AST-0216105.
Financial
support from DoE grant DE-FG52-06NA26217 is acknowledged.

\clearpage

\newpage
\begin{center}
\begin{table}[t]
\caption{Parameters of Runs}
\begin{tabular}{ccccc} \hline \hline \\
Run & $B/B_c$ & $\lambda_c / L$ & $\lambda_{\rm max}/L$
& $\lambda_c / \Delta x$  \\ \hline \\
$RH$ & 0.0 & - & - & - \\
$R1$ & 0.1 & 0.01 & 0.02 & 2.56 \\
$R2$ & 0.2 & 0.04 & 0.08 & 10.2 \\
$R4$ & 0.4 & 0.16 & 0.32 & 41.0 \\
$R6$ & 0.6 & 0.36 & 0.72 & 92.2 \\
\end{tabular}
\end{table}
\end{center}

\begin{center}
\begin{table}[h]
\caption{Volume averaged energies at $h/L=0.5$.}
\begin{tabular}{cccccccc} \hline \hline \\
Run & $b_0/b_c$ & $t/t_s$ & $(b^2 - b_0^2)/2$ & $\rho v^2/2$ & $e-e_0$ & total 
& $\delta b^2/b_0^2$ \\ \hline \\
$RH$ & - & 53 & - & 1.6 & 2.5 & 4.1 & - \\
$R1$ & 0.1 & 46 & 0.85 & 1.6 & 2.7 & 4.4 & 10.6 \\
$R2$ & 0.2 & 42 & 1.0 & 1.3 & 2.1 & 4.4 & 3.2 \\
$R4$ & 0.4 & 42 & 1.2 & 1.4 & 2.1 & 4.7 & 0.93 \\
$R6$ & 0.6 & 49 & 1.1 & 1.2 & 2.2 & 4.5 & 0.37 \\
\end{tabular}
\end{table}
\end{center}

\begin{figure}
\epsscale{0.8}
\plotone{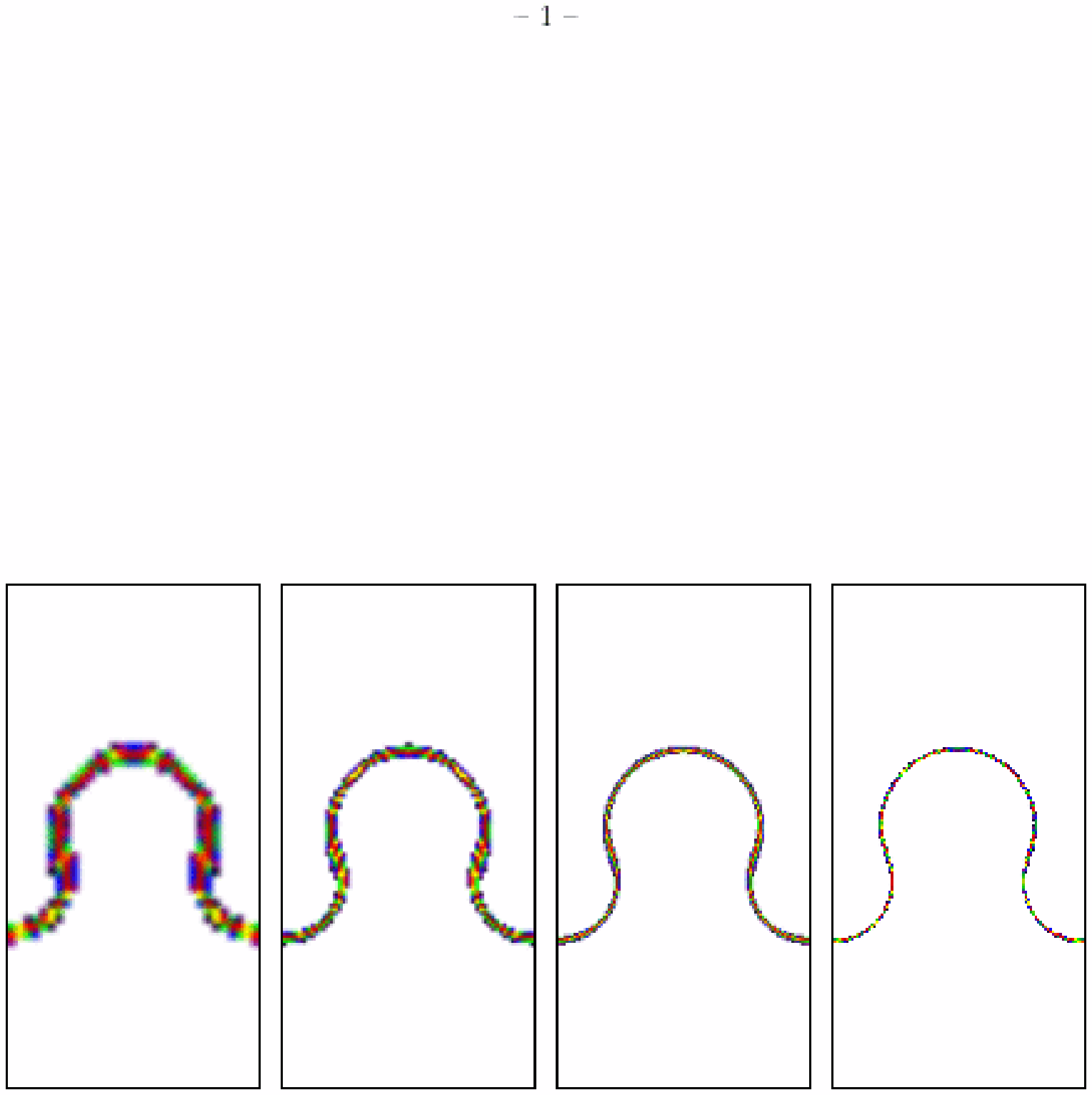}
\figcaption
{Images of the mixing parameter $\Theta$, defined in equation 12, in
the two-dimensional version of run R6 at $t/t_s=30.0$ at resolutions
of $32\times64$ (left), $64\times128$ (middle left), $128\times256$
(middle right), and $256\times512$ (right).  The color table runs blue to red
(online version) over the range zero to one,
with white corresponding to $\Theta=0$.} 
\end{figure}

\begin{figure}
\epsscale{0.8}
\plotone{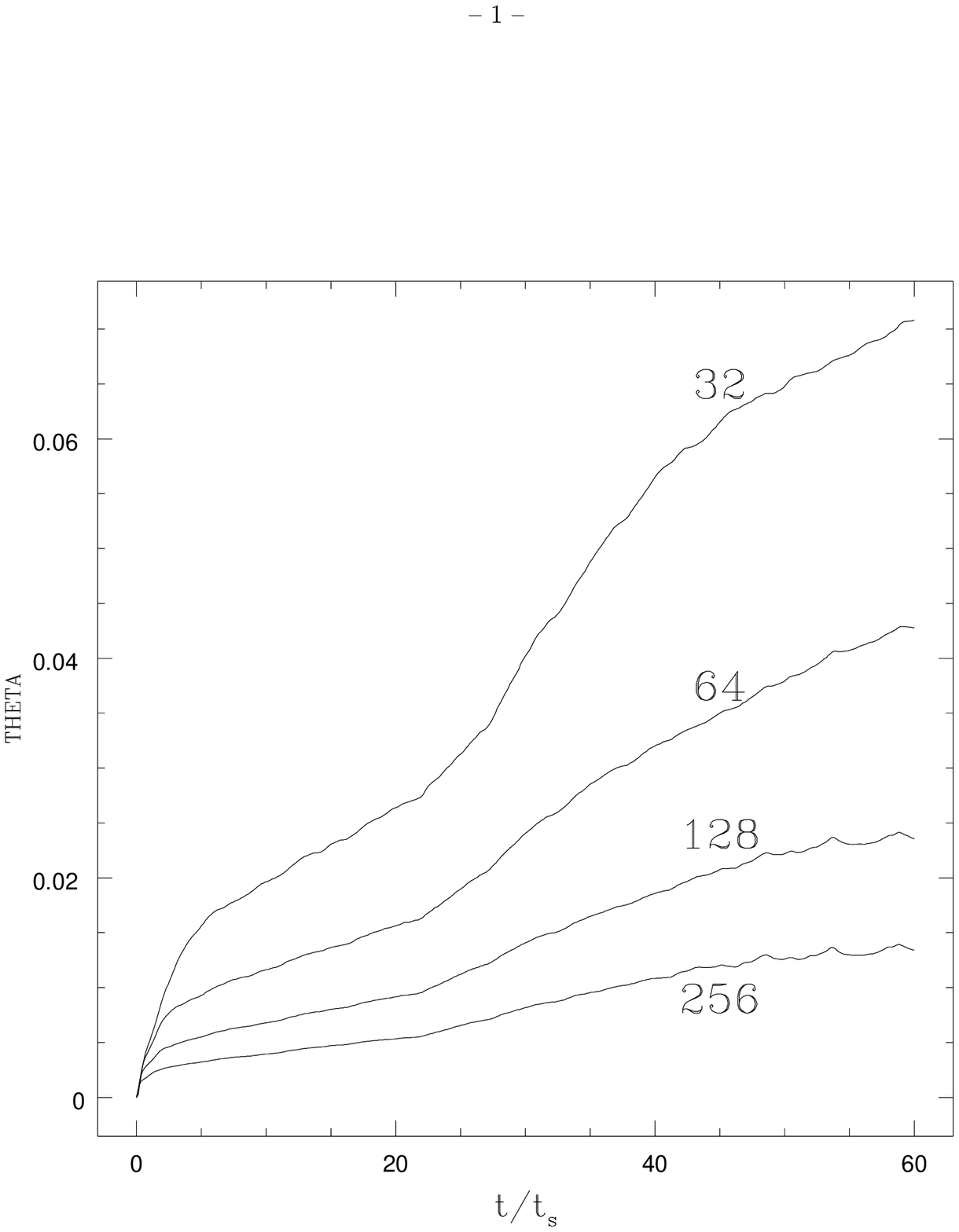}
\figcaption
{Time evolution of the volume-averaged mixing parameter at different 
resolutions for the two-dimensional version of the strong field run R6.
Each curve is labelled by the number of grid points per $L$.}
\end{figure}

\begin{figure}
\plotone{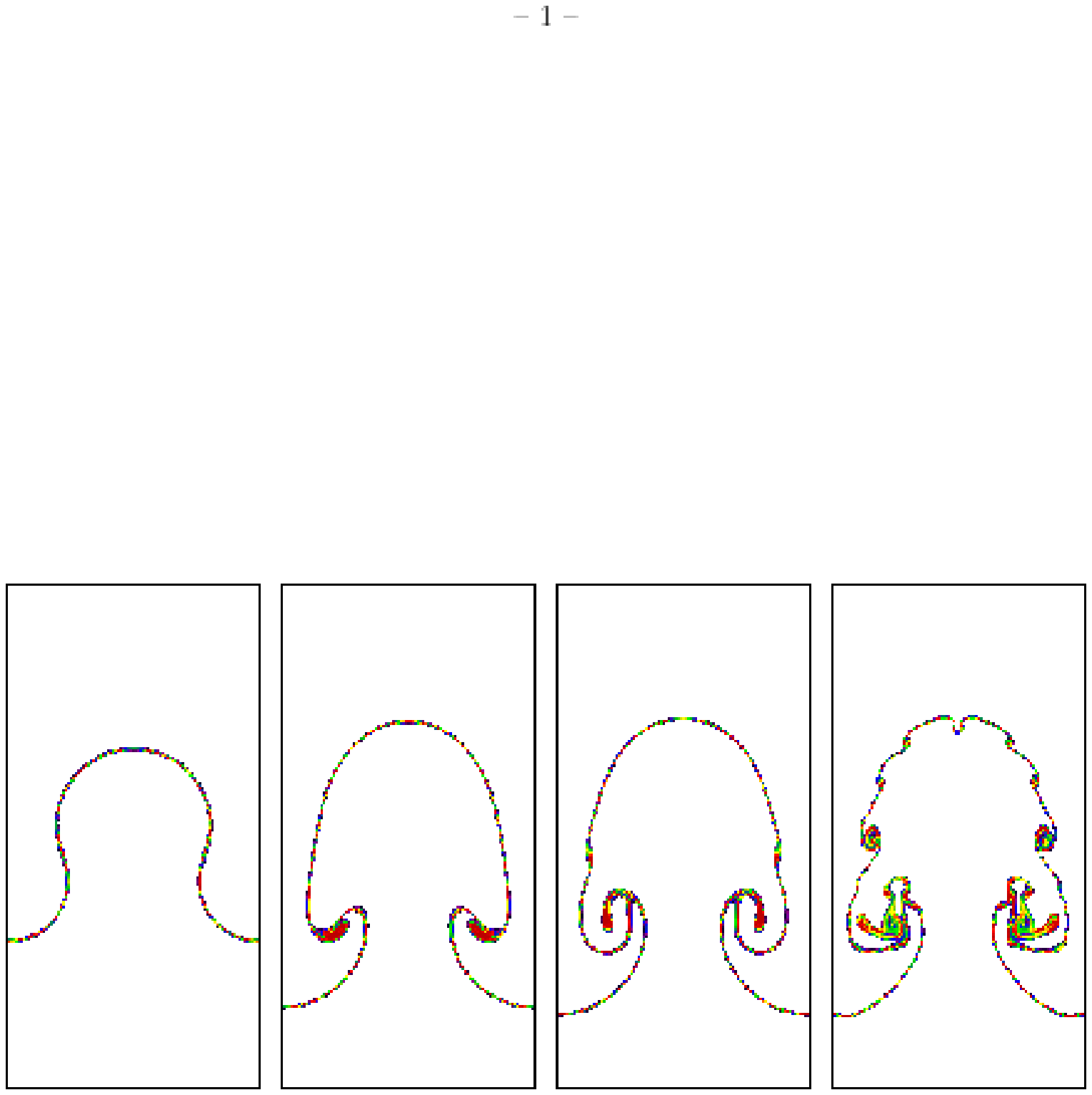}
\figcaption
{Images of the mixing parameter $\Theta$ at $t/t_s=30.0$, in
two-dimensional version of runs R6 (left, strong field), R2 (middle left,
intermediate field), R1 (middle right, weak field), and RH (right,
hydrodynamic), all at a resolution of $256\times512$. The
color table runs blue to red (online version) over the range zero to one, with white corresponding to $\Theta=0$.}
\end{figure}

\begin{figure}
\epsscale{0.8}
\plotone{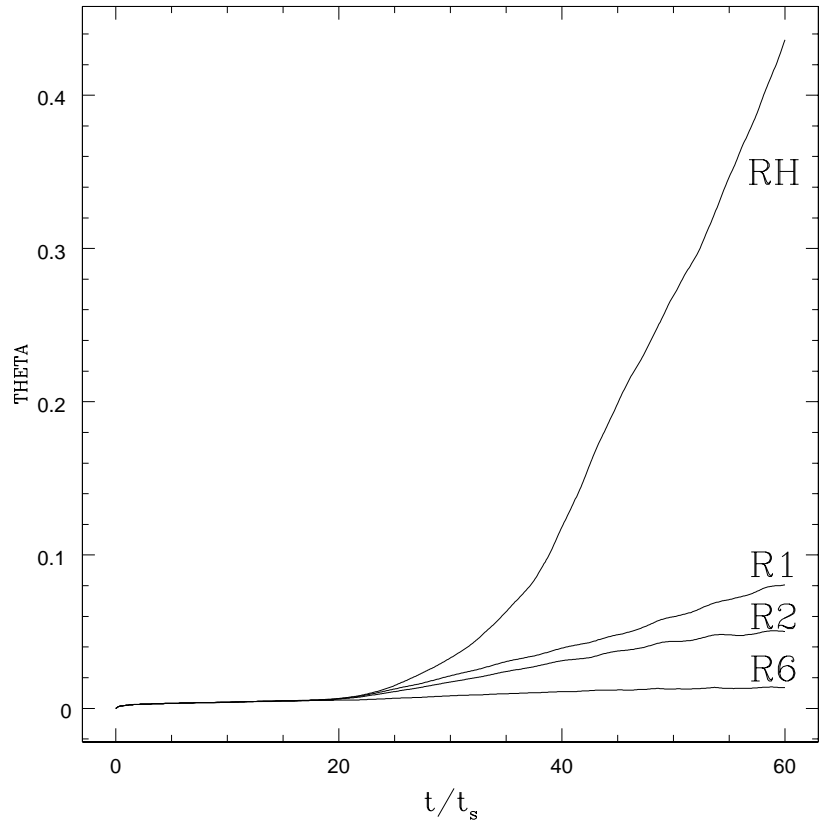}
\figcaption
{Time evolution of the volume-averaged mixing parameter for
runs R6 (strong field), R2 (intermediate field), R1 (weak field) and RH
(hydrodynamic).  The rapid increase in the mixing rate occurring at
$t/t_s=20$ in runs with weaker fields is due to the formation of KH rolls,
as is evident in figure 3.}
\end{figure}

\begin{figure}
\plotone{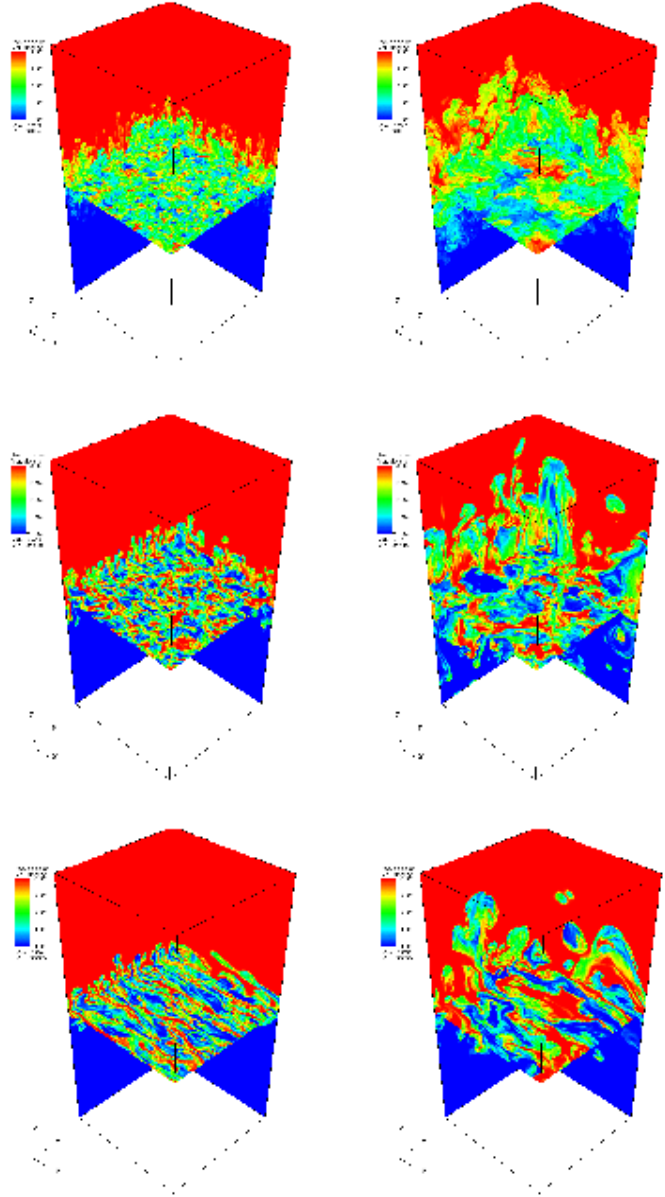}
\figcaption
{Slices of the density at $t/t_s = 29.6$ (left panels) and $t/t_s=60.0$
(right panels) in runs RH (top, hydrodynamic), R2 (middle, weak field),
and R6 (bottom, strong field).  Note the decrease in mixing in the MHD RTI
(as evidenced by reduced volume at intermediate densities, i.e. grey regions),
and the elongation
of structures along the magnetic field ($x-$axis) in the strong field case.
(Online version in color.)}
\end{figure}

\begin{figure}
\plotone{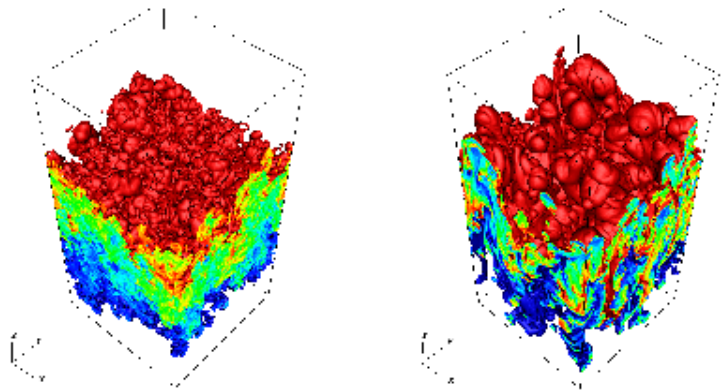}
\figcaption
{Isosurfaces of the density at $\rho=2.9$ and $\rho=1.1$
at $t/t_s=56.0$ in runs RH (left, hydrodynamic) and R2 (right, weak field).  Also shown are slices
of the density at the edges of the domain.
(Online version in color.)}
\end{figure}

\begin{figure}
\plotone{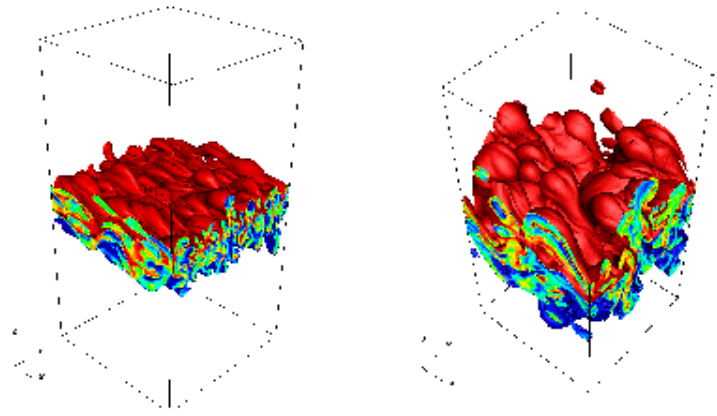}
\figcaption
{Isosurfaces of the density at $\rho=2.9$ and $\rho=1.1$
at $t/t_s=20.0$ (left) and $t/t_s=56.0$ (right) in run R6 (strong field).
Also shown are slices of the density at the edges of the domain.
(Online version in color.)}
\end{figure}

\begin{figure}
\epsscale{0.8}
\plotone{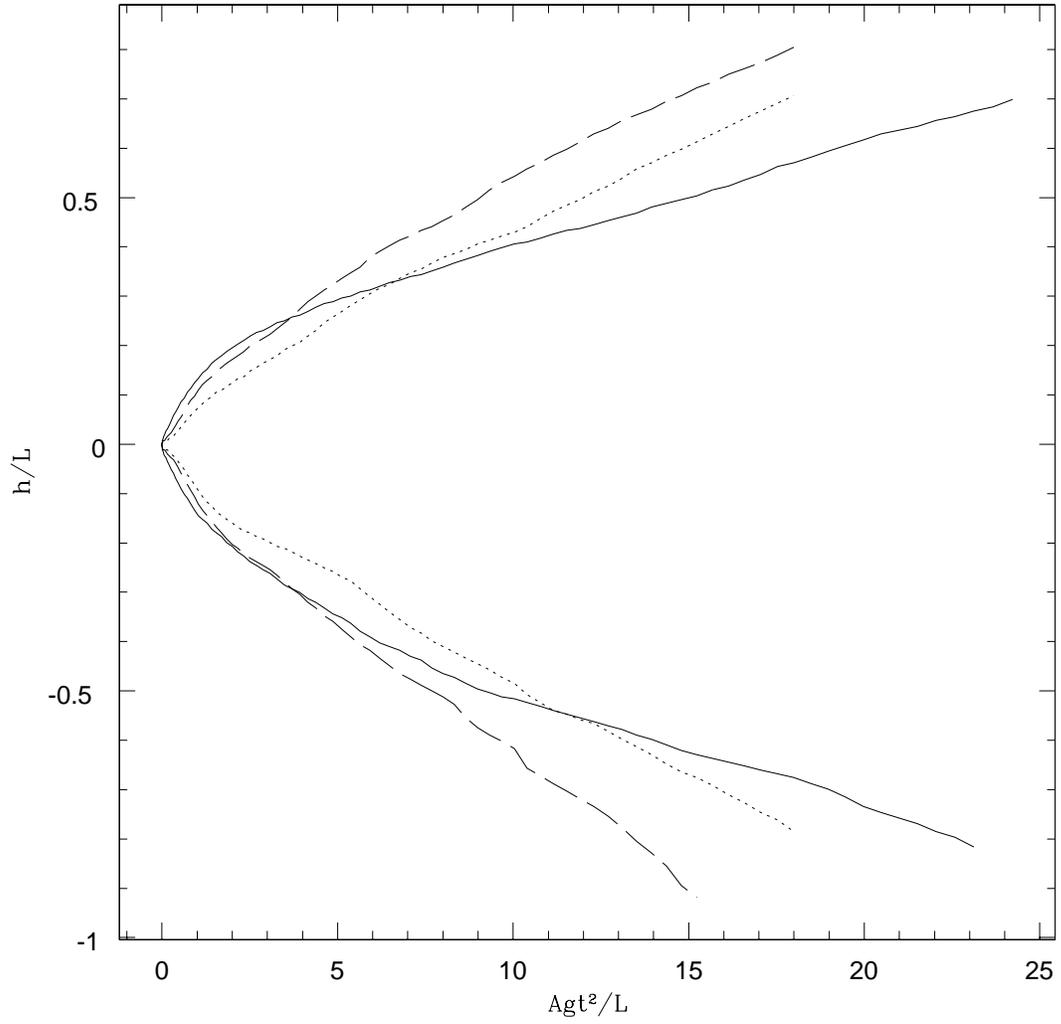}
\figcaption
{Height of bubbles ($z>0$) and fingers ($z<0$) as a function
of time in runs RH (solid line, hydrodynamic), R2 (dashed line, weak field)
and R6 (dotted line, strong field).
The height is defined as the location where the horizontally averaged
mass fraction (eq. 11) is 0.985 and 0.015 respectively.}
\end{figure}

\begin{figure}
\epsscale{0.8}
\plotone{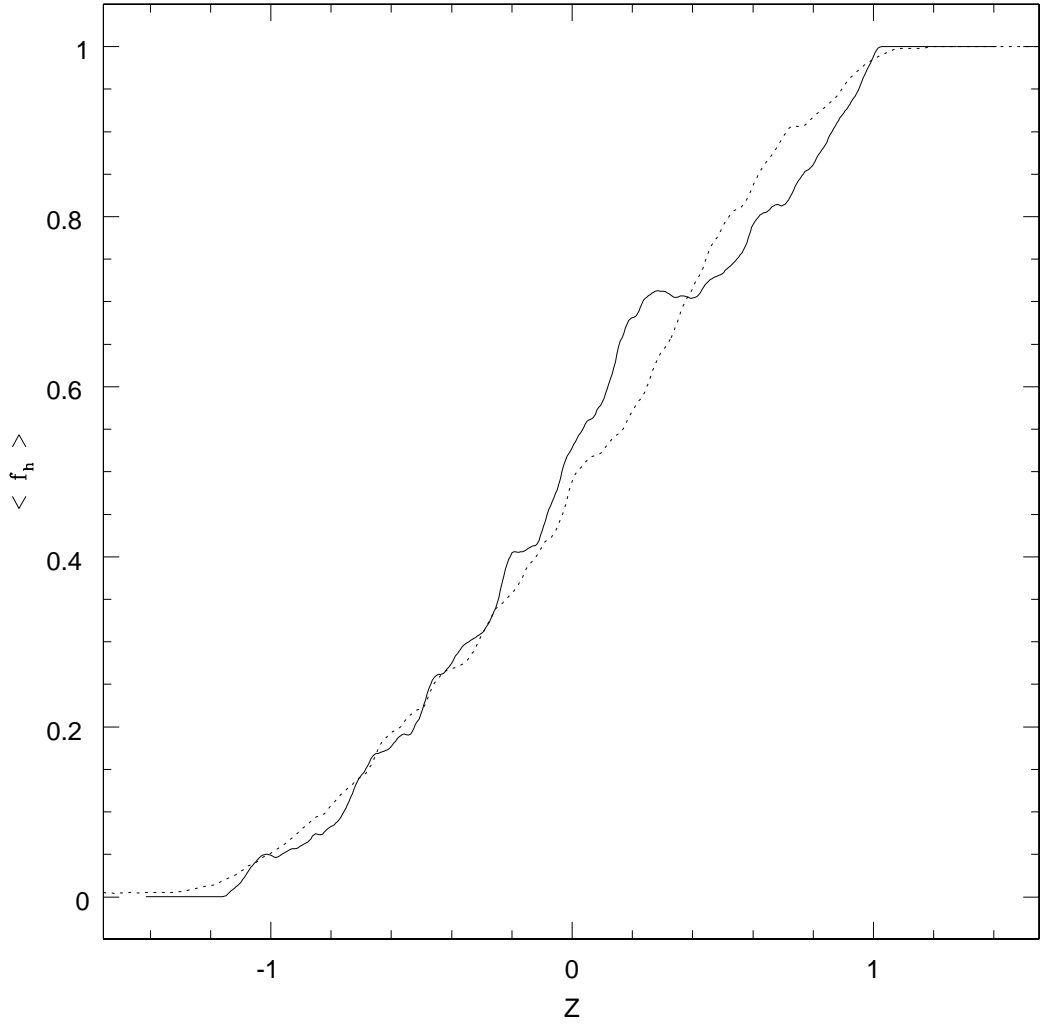}
\figcaption
{Vertical profile of the mass fraction $\langle f_h \rangle$ defined by eq. 11
in runs RH (dashed line, hydrodynamic) and R6
(solid line, strong field) at $t/t_s = 56$.  The horizontal axis has been scaled by the 
height of the bubbles at that time.}
\end{figure}

\begin{figure}
\epsscale{0.8}
\plotone{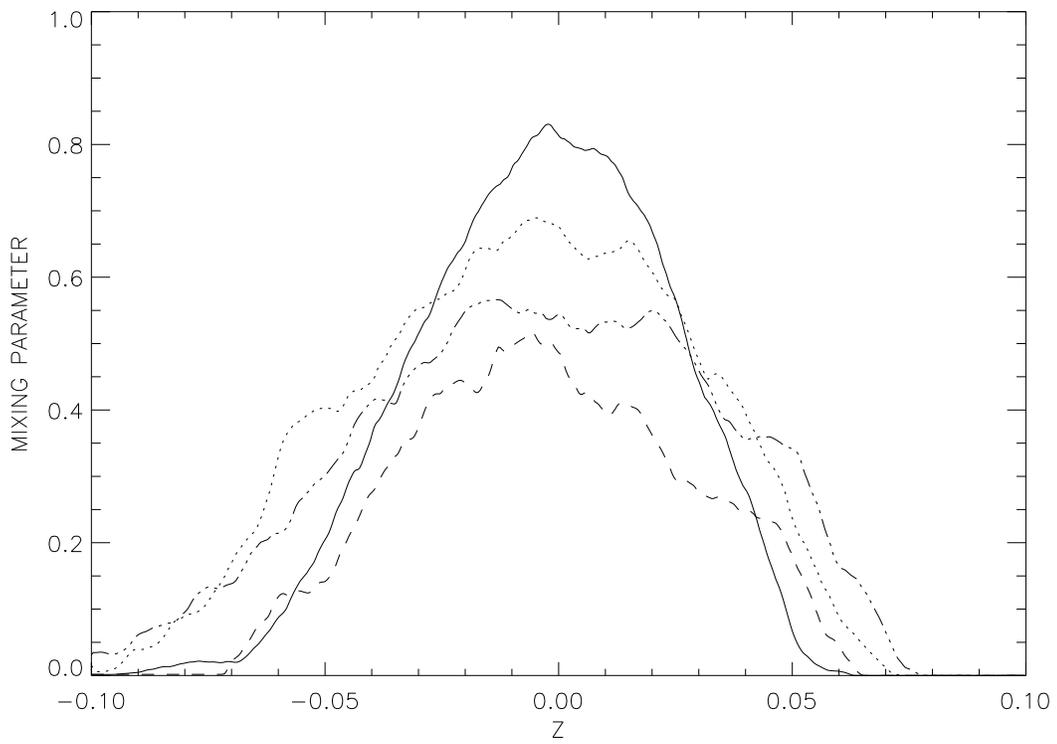}
\figcaption
{Vertical profile of the horizontally averaged mixing parameter $\langle \Theta \rangle \equiv 4 \langle f_h f_l
\rangle $ at time $t/t_s=56$ in runs RH (solid line, hydrodynamic),
R1 (dotted line, very weak field),
R2 (dot-dash line, weak field), and R6 (dashed line, strong field).
As measured by $\langle \Theta \rangle$,
there is monotonically less mixing with increasing field strength.}
\end{figure}

\begin{figure}
\epsscale{0.8}
\plotone{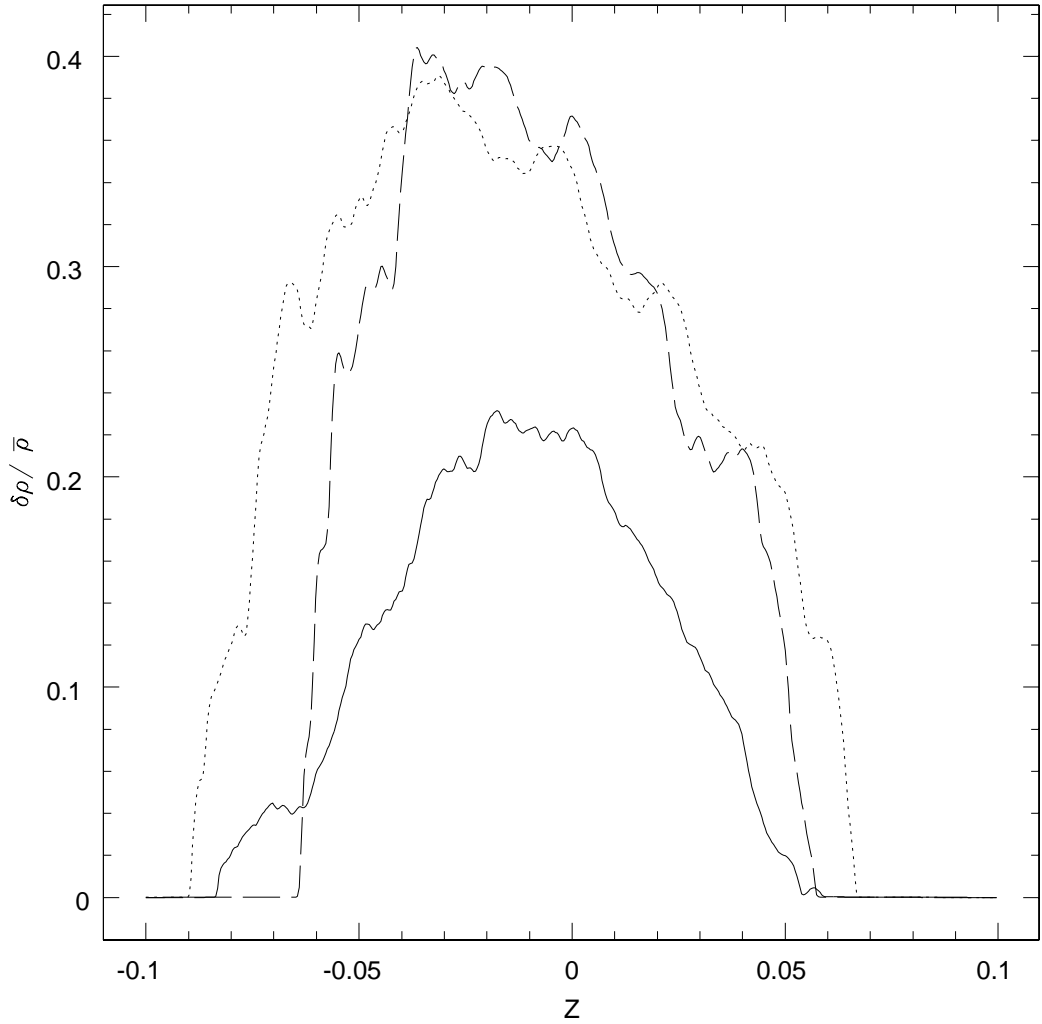}
\figcaption
{Vertical profile of the variance of the density defined by eq. 13 
in RH (solid line, hydrodynamic), R2 (dotted line, weak field) and R6 
(dashed line, strong field).  Both of the
MHD runs show a significantly larger variance than the hydrodynamic case
indicating less mixing.}
\end{figure}

\begin{figure}
\epsscale{0.8}
\plotone{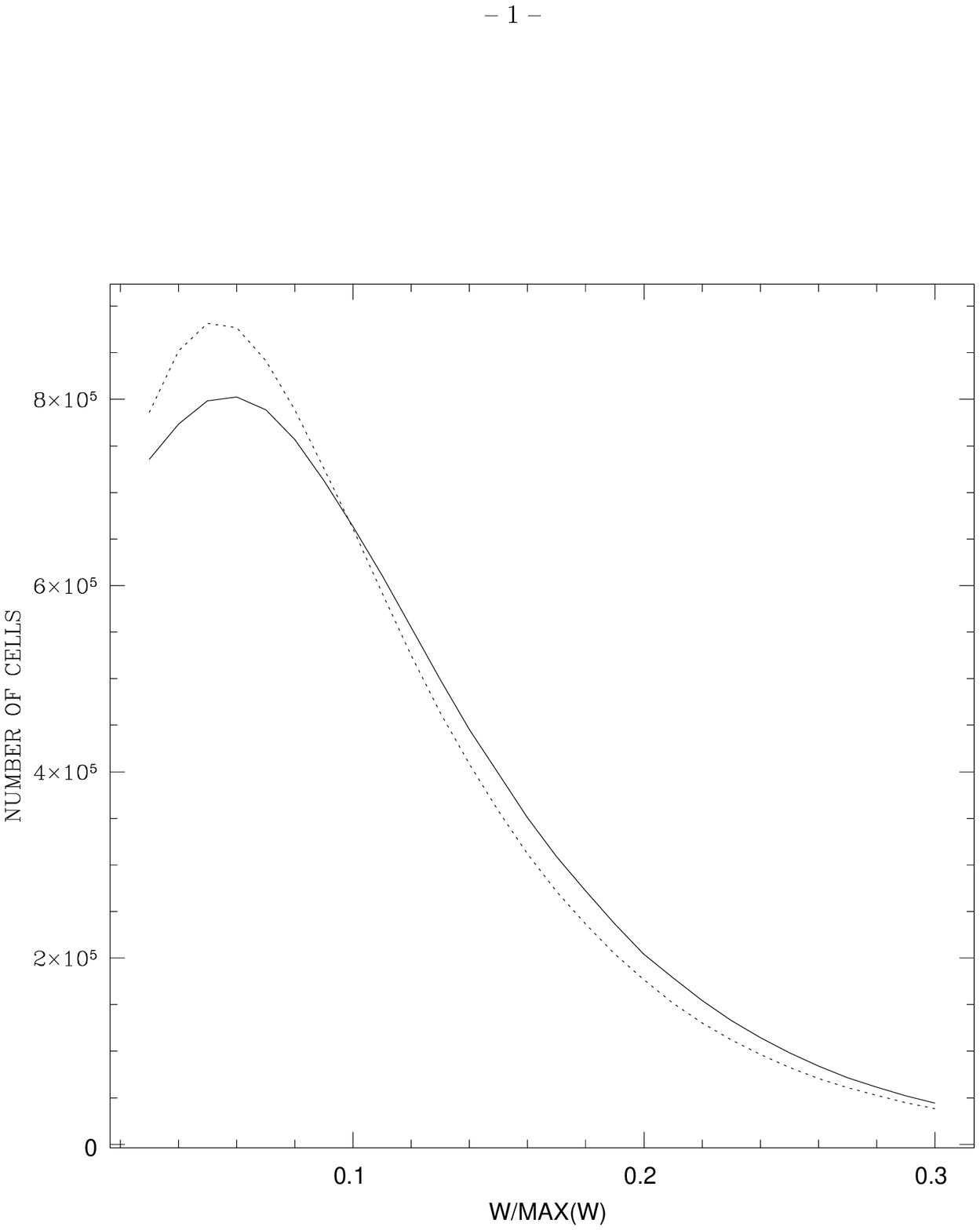}
\figcaption
{Distribution of the amplitude of the vorticity $W$ in runs RH (solid line,
hydrodynamic) and R1 (dotted line, very weak field) at
$t/t_s=40$.  The vertical axis is the number of cells with the given value
of $W$.  The horizontal axis is scaled by the maximum of $W$ in run RH.  }
\end{figure}

\begin{figure}
\plotone{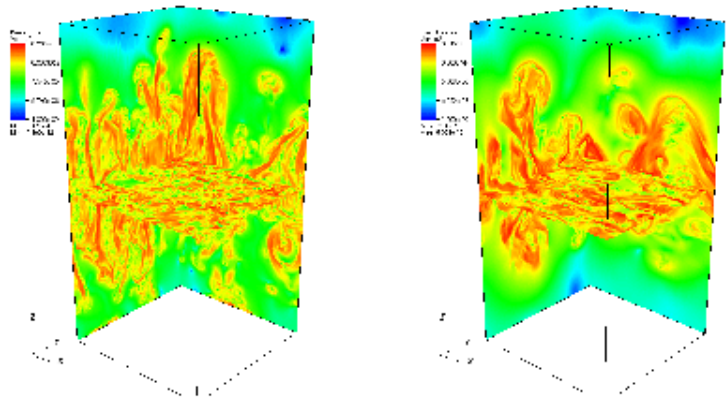}
\figcaption
{Slices of the magnetic energy in fluctuations, defined by eq. 14, 
at $t/t_s = 60$ in runs R2 (left, weak field) and R6 (right, strong
field).  The slices are taken at the same locations and at the same time
as the right-hand panels in figure 1.
(Online version in color.)}
\end{figure}

\begin{figure}
\plotone{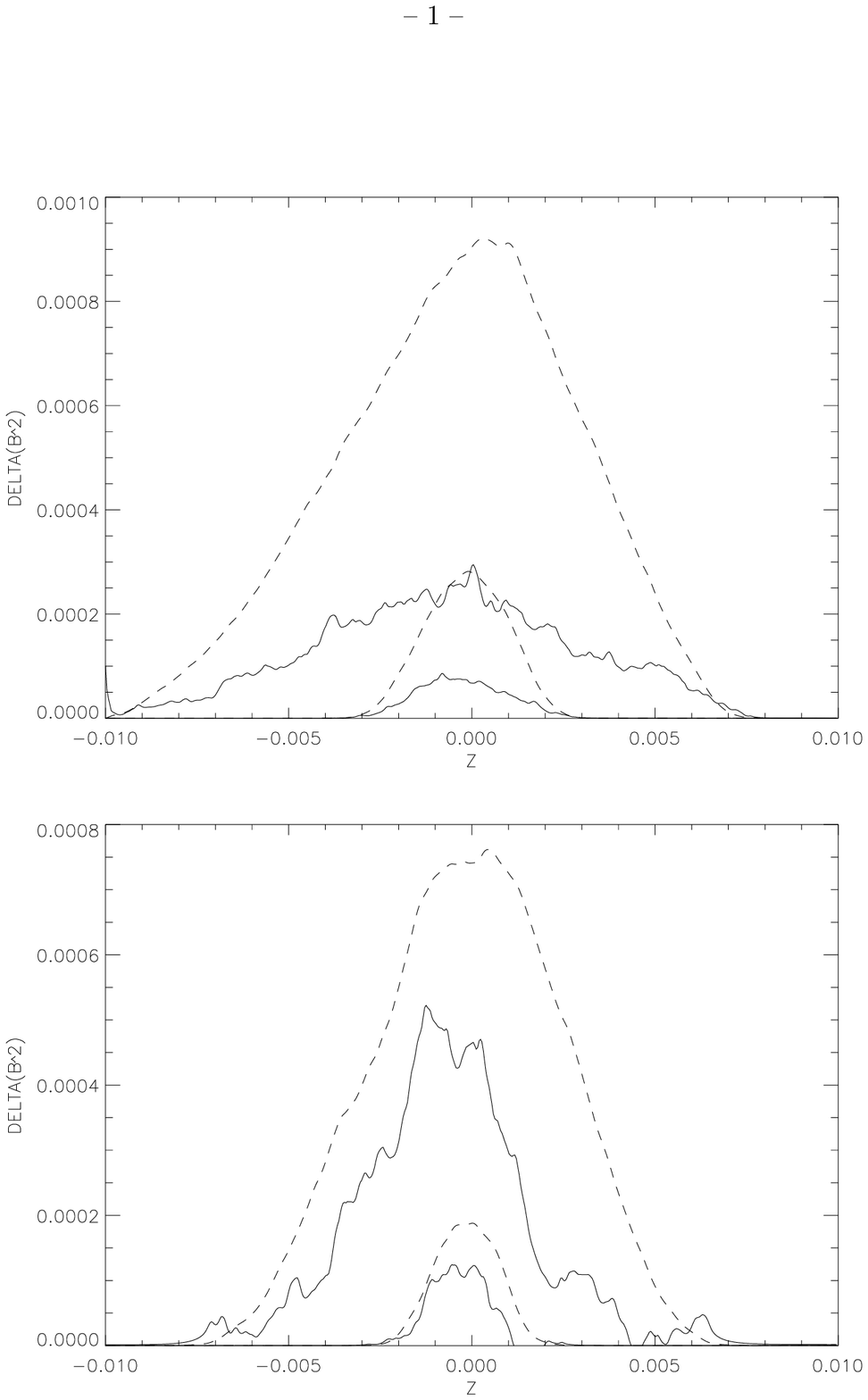}
\figcaption
{Profiles of the horizontally averaged magnetic energy in fluctuations for
runs R2 (top, weak field) and R6 (bottom, strong field).  The dashed lines
in each plot correspond to the energy in the vertical component of the field
$B_{z}^{2}$, while the solid lines correspond to the the energy in the
horizontal components of the field $(B_{x}-B_0)^{2} + B_{y}^{2}$.  The
profiles are shown at $t/t_s = 28$ and $t/t_s = 56$ in each plot, the
latter pair of lines extend over a larger horizontal range in both plots.}
\end{figure}

\begin{figure}
\plotone{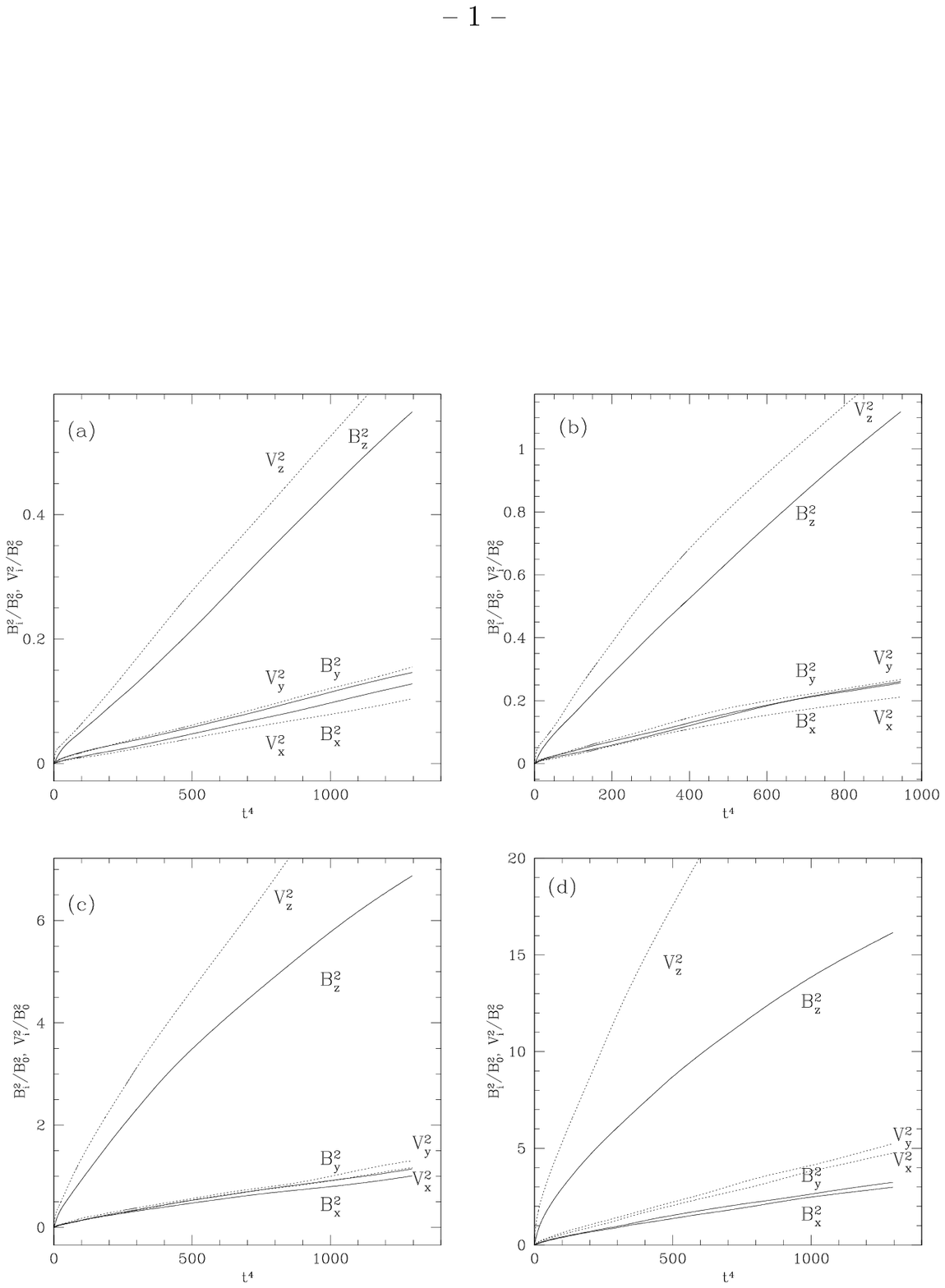}
\figcaption
{Time evolution of each component of the magnetic and kinetic energies in runs
{\em (a)} R6, strong field, {\em (b)} R4, intermediate field,
{\em (c)} R2, weak field, and {\em (d)} R1, very weak field.  Solid lines
show magnetic energy, dashed lines are kinetic.  Each line is labeled by the
associated field component.  The energy in perturbations is shown for
the $x-$component of the magnetic field, $\delta B_{x}^{2}/2 = (B_x^2 - B_{0}^{2})/2$, and in each panel the values are scaled by the initial magnetic energy
$B_{0}^{2}/2$.}
\end{figure}

\end{document}